\documentstyle[array,multicol,aps,preprint]{revtex}
\tightenlines

\begin{document}
\preprint{{hep-th/0410258} \hfill {UCVFC-DF-16-2004}}
\title{Magnetic Monopole in the Loop Representation}
\author{Lorenzo Leal{${}^{a }$}\footnote{lleal@fisica.ciens.ucv.ve} and
Alexander L\'{o}pez {${}^{b, \,a
}$}\footnote{alopez@ivic.ve}}
\address{${}^a$Grupo de Campos y Part\'{\i}culas, Departamento de F\'{\i}sica, Facultad de Ciencias, Universidad
Central de Venezuela, AP 47270, Caracas 1041-A, Venezuela \\
${}^b$Centro de F\'{\i}sica, Instituto Venezolano de
Investigaciones Cient\'{\i}ficas, AP 21827, Caracas 1020-A,
Venezuela\\}
%

\maketitle

\begin{abstract}
\noindent We quantize, within the Loop Representation formalism,
the electromagnetic field  in the presence of a static magnetic
pole. It is found that  loop-dependent physical wave functionals
acquire a topological dependence on the surfaces bounded by the
loop. This fact generalizes what occurs in ordinary quantum
mechanics in multiply connected spaces. When Dirac's quantization
condition is satisfied, the dependence on the surfaces disappears,
together with the influence of the monopole on the quantized
electromagnetic field.
\end{abstract}

\section{Introduction}
Dirac \cite{dirac,dirac2} found that the mere existence of a
single monopole would explain the quantized nature of the electric
charge. He discovered a relation between the unit of electric
charge and that of the magnetic pole, which is currently known as
Dirac's quantization condition. In rationalized units
($c=\hbar=1$) it reads
\begin{equation}\label{uno}
\frac {eg}{4\pi}=\frac {1}{2}n ,
\end{equation} where $n$ is an integer and $e$ ($g$) is the unit of electric (magnetic) charge.
Besides this remarkable prediction, the Dirac theory of magnetic
monopoles has been a source of inspiration for the development of
new ideas in theoretical physics. Often, it happens that different
approaches to understand  the formulation of Dirac bring out
novelties or unexpected relationships between old things.

 Being a gauge theory, the Dirac theory of magnetic poles should
be a candidate to admit a quantum geometric representation, such
as the Loop Representation (LR) of Maxwell theory
\cite{dibartolo,gambini1,gambini2}. In this article we address
this point to some extent. Concretely, we study the LR formulation
of quantum Maxwell theory in presence of an external magnetic
pole, taking as starting point a first order action due to
Schwinger \cite{schwinger},  which is based on the earlier Dirac
theory \cite{dirac,dirac2}. We shall see that the LR formulation
of the Maxwell theory with a static monopole corresponds to that
of the \emph{free theory}, except by the fact that the
loop-dependent wave functional acquires a topological dependence
on the manner that the loop ''winds'' around the monopole. This
dependence is manifested through a topological phase factor picked
up by the wave functional when the loop undergoes an adiabatic
excursion in the presence of the monopole. It could  be said that
loop-dependent wave functionals become multivalued in the presence
of the monopole, in the same sense that in ordinary quantum
mechanics wave functions are allowed to be multivalued whenever
the configuration space is multiply connected
\cite{morandi,balachandran,balachandran2}. This results  are also
related with previous studies about the quantum theory of  strings
in presence of a Kalb-Ramond vortex \cite {setaro,leal-ab}
 where a generalization of
the concept of anyons can be envisaged.

In the next section we present the model and discuss its
quantization in the LR formulation. In the last one we study in
which sense the magnetic monopole turns Maxwell theory into a
loop-dependent theory with non-trivial boundary conditions in
loop-space.

\section{Quantization and Loop Representation}\label{sec2}

 Electromagnetism with magnetic charges can be studied from the
first order Schwinger action \cite{schwinger}
\begin{equation}\label{dos}
S=\int dx^{4} \left (A_{\mu} J^{\mu}_e+B_{\mu} J^{\mu}_m-\frac
{1}{2}F^{\mu\nu}(\partial_{\mu}A_{\nu}-\partial_{\nu}A_{\mu})+\frac
{1}{4}F^{\mu\nu}F_{\mu\nu}\right ),
\end{equation}
where $B$ is given by
\begin{eqnarray}\label{tres}
B_{\mu}(x)&=& \int dy^{4}
*F_{\mu\nu}(y)f^{\nu}(y-x)+\partial_{\mu}\lambda(x).
\end{eqnarray}
Here,  $f$ obeys
\begin{eqnarray}
\partial_{\mu}f^{\mu}(y)&=&\delta^{4}({y}),\label{cuatro}
\end{eqnarray}
and $\lambda$ is an arbitrary function. The dual $*F$ is given by
\begin{equation}
*F_{\mu\nu}=\frac{1}{2}\epsilon_{\mu\nu\alpha\beta}F^{\alpha\beta},
\end{equation}
where $\epsilon_{\mu\nu\alpha\beta}$ is the completely
antisymmetric symbol and $J_e$ ($J_m$) denote the electric
(magnetic) current density. The independent fields in (\ref{dos})
are $A_{\nu}$ and $F_{\mu\nu}$.
 Varying (\ref{dos}) with respect to $A_{\mu}$ gives
\begin{eqnarray}
\partial_\nu F^{\mu\nu}&=&J^{\mu}_e,
\end{eqnarray}
whereas variations with respect to $F_{\mu\nu}$ yield
\begin{equation}
F_{\mu\nu}(x)=\partial_{\mu}A_{\nu}(x)-\partial_{\nu}A_{\mu}(x)-\frac{1}{2}\epsilon_{\mu\nu\alpha\beta}\int
dy^{4} J^{\alpha}_m(y)f^{\beta}(x-y),
\end{equation}
which, with the use of equations (\ref{tres}) and (\ref{cuatro}),
implies
\begin{eqnarray}
\partial_\nu *F^{\mu\nu}&=&J^{\mu}_m.
\end{eqnarray}
Thus, one obtains the Maxwell equations with both electric and
magnetic currents. The duality electricity-magnetism manifests
through the invariance of the equations under the rotations
\begin{eqnarray}
J_e &\rightarrow & \cos\phi J_e+\sin\phi J_m ,\\
J_m &\rightarrow & -\sin\phi J_e+\cos\phi J_m, \\
F &\rightarrow & \cos\phi F +\sin\phi *F.
\end{eqnarray}

  We are interested in studying how the presence of a magnetic
monopole affects the loop-space formulation of the Maxwell field.
Hence, we take $J_e=0$ in equation (\ref{dos}) and restrict
ourselves to consider a static monopole (we take it at the origin
of space), which forces the magnetic current to be written as
\begin{equation}
J^{\mu}_m(x)=g\delta^{\mu}_0 \delta^{3}(\vec{x}).
\end{equation}
A convenient choice for $f_{\mu}$ (fulfilling equation
(\ref{cuatro})) is
\begin{equation}\label{f}
f^{\nu}(y)=-\frac{1}{4\pi}\frac{y^{i}}{\vec{y}\,^3}\delta^{\nu}_{i}
\delta(y_0).
\end{equation}
 The Hamiltonian formulation begins with the definition of the canonical
 momenta associated to  $A_0$,  $F_{ij}$ and  $A_i$, which result to be
\begin{eqnarray}
\label{siete2}\Pi^{0}(x)&\approx &0,\\
\label{ocho2} \Pi^{ij}(x)&\approx&0, \\
\Pi^{i}(x)&=&-F^{0i}(x),
\end{eqnarray}
respectively. Equations (\ref{siete2}), (\ref{ocho2}) are
(primary) constraints in the sense of Dirac \cite{dirac3}, and we
have introduced the weak equality symbol $\approx $ to mean that
these equalities should not be used until Poisson brackets are
calculated. Since $F^{0i}$ is already a conjugate momentum, it is
not necessary to treat it as a coordinate and to define its own
canonical momentum \cite{jackiw}. Following  Dirac's method to
deal with constrained systems one constructs the total Hamiltonian
\cite{dirac3}
\begin{equation}\label{hamilton}
H^{*}= H +\int d\vec x\,^{3} \, u(x)\Pi^0(x)+\int d\vec x\,^{3} \,
u_{ij}(x)\Pi^{ij}(x),
\end{equation}
where $u(x)$ and $u_{ij}(x)$ are Lagrange multipliers and $H$ is
the canonical Hamiltonian
\begin{equation}\label{canon}
H=\int d\vec x\,
^{3}\left[\frac{1}{2}\Pi_{i}^2-\frac{1}{4}F_{ij}^2+\frac{1}{2}F_{ij}\left(f_{ij}-b_{ij}\right)\right]+\int
d \vec x\,^{3}\partial_i\Pi^{i} A_0.
\end{equation}
Here we have  defined
\begin{eqnarray}\label{byf}
f_{ij}(x)&=& \partial_i A_j(x)-\partial_j A_i(x),\\
b_{ij}(x)&=& g\epsilon_{ijk}f^k(\vec x).
\end{eqnarray}
It is worth noticing that  when the static monopole is absent this
Hamiltonian properly reduces to that of the conventional Maxwell
theory, since in that case the magnetic field $F_{ij}$ and the
curl of the potential $f_{ij}$ coincide.

The non-vanishing equal time canonical Poisson brackets are given
by
\begin{eqnarray}
   \{F_{ij}(x),\Pi^{kl}(y)\}&=&  \frac
12\left(\delta_{i}^{k}\delta_{j}^{l} -\delta_{i}^{l}
\delta_{j}^{k}\right)\delta^{3}(\vec{x}-\vec{y}). \\
   \{A_\alpha(x),\Pi^\beta(y)\}&=&\delta^\alpha_\beta \delta^{3}(\vec{x}-\vec{y}).
\end{eqnarray}
Following Dirac \cite{dirac3}, one must impose the time
preservation of the constraints. From (\ref{siete2}) one finds, as
a secondary constraint, the Gauss´ Law
\begin{equation}\label{gauss}
\partial_i\Pi^i(x)\approx 0.
\end{equation}
This constraint, in turn, has vanishing Poisson bracket with the
total Hamiltonian, hence, it does
 not produce further constraints.
On the other hand, time preservation of (\ref{ocho2}) gives the
secondary constraint
\begin{equation}\label{k}
K_{ij}(x)=F_{ij}(x)-f_{ij}(x)+b_{ij}(x)\approx 0.
\end{equation}
Finally, the  preservation of constraint (\ref{k}) allows to
obtain the Lagrange multipliers $u_{ij}$ in terms of the momenta
$\Pi^i$
\begin{equation}\label{u}
u_{ij}(x) = -\frac{1}{2}\left(\partial_i F_{0j}(x)-\partial_j
F_{0i}(x)\right),
\end{equation}
which  must be substituted into the expression for the total
Hamiltonian.

  It is found that constraints (\ref{gauss})
and (\ref{siete2}) are first class, while (\ref{ocho2}) and
(\ref{k}) are second class. Following Dirac's procedure we must
introduce  Dirac brackets in order to obtain a quantum theory
consistent with these second class constraints. Though it is not
difficult to carry out the calculations of the matrix of the
Poisson brackets between second class constraints and its inverse,
which are the ingredients needed for building Dirac brackets, a
simple argument suffices to get the result. Since Dirac brackets
are going to be consistent with  second class constraints, these
may be put as strong equalities. Hence, we can write $F_{ij}$ and
its momentum $\Pi^{ij}$ in terms of the remaining canonical
variables, and substitute these expressions in the total
Hamiltonian. Once this is done, we have only to consider  Dirac
brackets between the canonical variables that remain, which are
$A_i$, $A_0$, and their canonical conjugates. But it is easy to
see that these Dirac brackets just coincide with the Poisson ones.
The net result is that we can eliminate  $F_{ij}$ and $\Pi^{ij}$
using equations (\ref{ocho2}) and (\ref{k}), and continue using
the Poisson brackets for the remaining variables.

At this point, it is also convenient to "eliminate" the constraint
(\ref{siete2}) of the formalism. This can be accomplished by
fixing the temporal gauge $A_{0}=0$, and treating  this equation
as a new constraint, which, together with (\ref{siete2}), can be
considered as a pair of second class constraints.  Then, we can
put $A_{0}$ and $\Pi^{0}$ as strongly vanishing. As before, it can
be seen that the new Dirac brackets are equal to the Poisson ones,
as far as we consider only the remaining variables, namely $A_i$
and their canonical conjugates.

Now we are ready  to  quantize  the theory.  First, promote
canonical variables to operators obeying equal time canonical
commutators
\begin{eqnarray}
\label{algebra1} [\hat{A}_i(x),\hat{A}_j(y)]&=& 0,\\
\label{algebra2} \left[\hat{\Pi}^i(x),\hat{\Pi}^j(y)\right]&=& 0,\\
\label{algebra3} \left[ \hat{A}_i(x),\hat{\Pi^j}(y)\right]&=&
i\delta^{j}_i\delta^{3}(\vec x-\vec y).
\end{eqnarray}
 The first class constraints define the physical states $|\Psi>$ as
those that satisfy
\begin{eqnarray}\label{c1}
\partial_i\hat{\Pi}^i(x)|\Psi>&=& 0.  \label{c2}
\end{eqnarray}
On the physical subspace, the dynamics is given by the
Schr\"{o}dinger equation
\begin{equation} i\partial_t|\Psi_t>=\hat{H}|\Psi_t>,
\end{equation}
with
\begin{equation}\label{phys}
\hat{H}=\int d\vec x\,^{3}\left[\frac{1}{2}\hat{\Pi}
_{i}^2+\frac{1}{4}\left(\partial_i \hat{A}_j-\partial_j
\hat{A}_i-b_{ij}\right)^{2}\right].
\end{equation}
Thus, we obtain that the static monopole manifests in the theory
just through the external field $b_{ij}$, that must be subtracted
from the curl of the vector potential to give the magnetic field
operator.

 We  are  now prepared to discuss the LR
of the model. We begin by recalling that the Abelian path space
(PS) can be defined as the set of certain equivalence classes of
curves $\gamma$ in (for our purposes) $R^{3}$
\cite{dibartolo,gambini1,gambini2,camacaro}. The equivalence
relation is given by the so called { \emph{form factor}  $T^i(\vec
x,\gamma)$ of the curve
\begin{equation}
T^i(\vec x,\gamma)=\int_\gamma dy^i\delta(\vec x-\vec y),
\end{equation}
as follows: $\gamma$ and $\gamma ^{,}$ are said to be equivalent
(i.e., represent the same path) if their form factors coincide.
Closed curves give raise to a subspace of the PS: the loop space.
It can be seen that the  usual composition of curves translates
into a composition of paths that endows the PS with a group
structure.

The path representation arises when one considers path dependent
wave functionals $\Psi[\gamma]$, and realizes the canonical
 field operators by means of operations onto these wave
 functionals \cite{dibartolo,gambini1,gambini2,camacaro}. We define the path and loop
 derivatives  $\delta_i(\vec x)$ and  $\Delta_{ij}(\vec x)$ by
\begin{eqnarray}
\left(1+u^i(\vec x){\delta}_{i}(\vec
x)\right)\Psi[\gamma]&=&\Psi[\gamma\circ u],\\
\left(1+\frac{1}{2}\sigma^{ij}(\vec x){\Delta}_{ij}(\vec
x)\right)\Psi[\gamma]&=&\Psi[\gamma\circ\delta c],
\end{eqnarray}
where $\circ$ denotes the PS product \cite{gambini1}. The
derivative $\delta_i(\vec x)$ ($\Delta_{ij}(\vec x)$)  measures
the change in the path-dependent wave functional   when an
infinitesimal path $\delta u$ (infinitesimal loop $\delta c$) is
attached to its argument $\gamma$ at the point $\vec x$. It is
understood that these changes are considered up to  first order in
the infinitesimal vector $u^i$ associated with the small path, or
with the surface element
\begin{equation}\label{2.18}
\sigma^{ij}=u^iv^j-v^ju^i ,
\end{equation}
generated by the infinitesimal vectors $\vec{u}$ and $\vec{v}$
that define the small loop $\delta c$. It can be shown that both
derivatives are related by
\cite{dibartolo,gambini1,gambini2,camacaro}
\begin{equation}
\partial_i{\delta}_j(\vec x)-\partial_j{\delta}_i(\vec x)={\Delta}_{ij}(\vec
x).
\end{equation}

 With these tools at hand we represent the canonical field as
 operators acting on path dependent wave functionals
 $\Psi[\gamma]$ by means of the prescriptions
\begin{eqnarray}
\hat{\Pi}^i(\vec x)&\rightarrow &e{T}^i(\vec x,\gamma),\\
\hat{A}_j(\vec x)&\rightarrow &\frac{i}{e}{\delta}_j(\vec x).
\end{eqnarray}
It is readily seen that this realizes  the algebra
(\ref{algebra1})- (\ref{algebra3}). We see that in this
representation the form factor corresponds to Faraday lines of
electric field. The magnetic field operator, in turn, appends a
small closed line of electric field  to the argument of the wave
functionals. The constant $e$ is introduced to fix the scale of
the  Faraday´s lines of electric field. In four dimensions (and
using natural units) this constant is dimensionless (as well as
the magnetic charge $g$) and can be taken as the elementary
electric charge. Since the divergence of the form factor $T^i(\vec
x,\gamma)$ vanishes when the path is closed, the Gauss constraint
(\ref{c1}) is identically satisfied if we restrict to
loop-dependent wave functionals \cite{dibartolo}.
 The Hamiltonian in the LR is then given by
\begin{equation}\label{phys2}
H =\int d\vec x\left[\frac{1}{2}e^2\left({T}^i(\vec
x,\gamma)\right)^2-\frac{1}{4e^2}\left(i{\Delta}_{ij}(\vec x)-
e\,b_{ij}(\vec x)\right)^2\right].
\end{equation}
When $g=0$, $H$ reduces to the Hamiltonian of free
electromagnetism in the LR \cite{dibartolo,gambini1,gambini2}, as
it should be.

\section{Multivalued Loop-dependent wave functionals}\label{sec3}

  We have seen that introducing  a static monopole in  quantum
  Maxwell theory, in the LR, amounts to replacing the loop derivative ${\Delta}_{ij}(\vec x)$
by a kind of ''covariant derivative''
\begin{equation}\label{mag}
i{\Delta}_{ij}(\vec x)  \rightarrow i{\Delta}_{ij}(\vec
x)-e\,b_{ij}(\vec x).
\end{equation}
Now, we shall see that  it is possible to recast the
Schr\"{o}dinger equation as that corresponding to a \emph{free}
theory, provided that we deal with \emph{multivalued
loop-dependent} wave functionals. To see how this happens we find
it convenient to employ the space of surfaces framework
\cite{pio1}, which we summarize following very closely reference
\cite{pio}. One starts with the space of piecewise smooth oriented
surfaces $\Sigma$ in $R^{3}$.   We define two surfaces as
equivalent if they
 share the same  "surface form-factor"
\begin{equation}\label{2.16}
T^{ij}(\vec{x},\Sigma)=\int
d\Sigma^{ij}_y\,\delta^{(3)}(\vec{x}-\vec{y}).
\end{equation}
Here $d\Sigma^{ij}_y$ is the surface element
\begin{equation}\label{2.18}
d\Sigma^{ij}_y=(\frac{\partial y^i }{\partial s}\frac{\partial
y^j}{\partial r} - \frac{\partial y^i}{\partial r}\frac{\partial
y^j}{\partial s})ds dr,
\end{equation}
with $s,r$ being surface parameters. Now we consider functionals
$\Psi[\Sigma]$ and introduce the surface derivative
$\delta_{ij}(\vec x)$, that measures the response of
$\Psi[\Sigma]$ when an element of surface whose infinitesimal area
is $\sigma_{ij}$ is attached to the argument $\Sigma$ of
$\Psi[\Sigma]$ at the point $x$, up to first order in
$\sigma_{ij}$ :
\begin{equation}\label{2.17}
\Psi [\delta\Sigma\cdot\Sigma]= (1 + \sigma^{ij}\delta_{ij}(\vec
x))\Psi [\Sigma].
\end{equation}
The surface derivative $\delta_{ij}(\vec x)$ and the loop
derivative $\Delta_{ij}(\vec x)$ are different things. However,
since in $R^{3}$ loop-dependence is a particular case of
surface-dependence (a loop is always the boundary of an open
surface in $R^{3}$), the loop derivative can be seen as the
surface derivative restricted to loop-dependent functionals. Hence
it makes sense to surface-derive loop-dependent quantities. Soon
we shall make use of the surface-derivative of the form factor
\begin{equation}\label{2.22}
\delta_{ij} (\vec{x}) T^{kl}(\vec{y},\Sigma)=\frac
12\left(\delta_{i}^{k}\delta_{j}^{l} -\delta_{i}^{l}
\delta_{j}^{k}\right)\delta^{(3)}(\vec{x}-\vec{y}).
\end{equation}

Turning back to our model, let us consider an open surface
$\Sigma$ whose boundary coincides with $\gamma$. Then define, from
the path-dependent wave functional $\Psi[\gamma]$, the
surface-dependent one
\begin{eqnarray}\label{sigma}
\Psi[\Sigma] &\equiv& \exp\left(ie\int d\Sigma^{km}_{\vec y}b
_{km}(\vec y) \right)\Psi[\gamma] \\
&=& \exp (\frac{ieg}{4\pi}\Omega(\Sigma))\,\Psi[\gamma] ,
\end{eqnarray}
where $\Omega(\Sigma)$ is the solid angle subtended by  $\Sigma$,
measured from the monopole.  Using equation (\ref{2.22}), it is
easy to show that
\begin{eqnarray}
\label{fin1}\delta_{ij}(\vec x)\Psi[\Sigma]=\exp (\frac{ieg}
{4\pi}\Omega(\Sigma))\left(\Delta_{ij}(\vec x)+i eb_{ij}(\vec
x)\right)\Psi[\gamma].
\end{eqnarray}
Then, the Schr\"{o}dinger equation of the theory can be written
down as
\begin{equation}\label{phys2}
i\partial_t \,\Psi[\Sigma ,t] =\int d\vec x\,
^{3}\left[\frac{1}{2}e^2\left({T}^i(\vec
x,\gamma)\right)^2-\frac{1}{4e^2}\left({\delta}_{ij}(\vec
x)\right)^2\right]\Psi[\Sigma ,t],
\end{equation}
which looks like the the Schr\"{o}dinger equation of the free
(i.e., without  charges or monopoles) Maxwell theory in the LR
\cite{gambini1,gambini2}, except by the presence of the surface
derivative instead of the loop derivative, and the surface
dependence (instead of simple loop-dependence) of the wave
functional. But since the only  property of $\Sigma$ that matters
is the solid angle subtended from the monopole, this
surface-dependence is topological: if we replace  $\Sigma$ by
another surface $\Sigma '$ that has the same boundary $\gamma$,
the wave functional changes as
\begin{equation}\label{topologica}
\Psi[\Sigma \,']=  \exp(i\,egp) \Psi[\Sigma],
\end{equation}
where $p$ is the number of times that the closed surface $ S=
\Sigma \,'\circ (-\Sigma)$, that results from the composition of
$\Sigma \,'$ and the surface opposite to $\Sigma$, wraps around
the monopole. Thus, we can take the Schr\"{o}dinger equation of
the Maxwell theory with external monopole as  that corresponding
to the free theory, provided that simultaneously we allow for
non-trivial boundary conditions for the loop-dependent wave
functionals: every time that the loop goes around a "closed
trajectory" that encloses $p$ times the monopole, the wave
function picks up the phase factor $\exp(i\,egp)$. The
loop-dependent wave functional becomes multivalued due to the
presence of the magnetic monopole, and can be finally written down
as a free loop-equation

\begin{equation}\label{phys2}
i\partial_t \,\Psi[\gamma ,t] =\int d\vec x\,
^{3}\left[\frac{1}{2}e^2\left({T}^i(\vec
x,\gamma)\right)^2-\frac{1}{4e^2}\left({\Delta}_{ij}(\vec
x)\right)^2\right]\Psi[\gamma ,t],
\end{equation}
where the multivalued  wave functional $\Psi[\gamma ,t]$ obeys the
boundary condition

\begin{equation}\label{topologica2}
\Psi[[S] . \gamma]=  \exp(i\,egp) \Psi[\gamma].
\end{equation}
In this equation, $[S].\gamma$ means that the loop $\gamma$ has
described a "closed trajectory"  sweeping  the surface $S$ and
wrapping $p$ times to the monopole.

 This can be understood as a generalization of what happens
in ordinary quantum mechanics in multiply connected configuration
spaces \cite{morandi,balachandran,balachandran2}. In such cases,
multivaluedness of the wave function is allowed, being restricted
to multiplication of the wave function by a phase factor carrying
a representation of the fundamental group of the configuration
space. But this is precisely what we have found in our study:
since the configuration space of our quantum formulation is the
space of loops  in $R^{3}- \{origin\}$, a ''point'' in the set is
a loop, while the ''closed curves'' swept by loops will be closed
surfaces, whose properties of contractibility in $R^{3}-
\{origin\}$ will define the fundamental group of the configuration
space. Now, the phase factor $ \exp(i\,egp)$ appearing in equation
(\ref{topologica}) just corresponds to a one-dimensional
representation of this fundamental group, since it classifies the
surfaces according to the manner they  wrap the monopole.

To conclude, we should mention a feature that could look somewhat
striking at first sight. If Dirac's quantization condition
(\ref{uno}) holds, the topological phase factor appearing in
(\ref{topologica}) becomes the unity and the dependence on the
surface vanishes. But then the wave functional becomes
single-valued and the effect of the monopole disappears at all!.
What happens is that, in the absence of  electrically charged
particles, there is no need to quantize  electric or magnetic
charges. Recall that in the formulation of Dirac, charge
quantization arises when the wave function of the  \emph{charged
particle} is asked to be single-valued in presence of the
monopole. Or, alternatively, when the action functional is asked
to be independent of the string attached to the monopole [in our
case, this corresponds to demanding that changing  $f^{\mu}$
(given in (\ref{f})) by any other vector field obeying equation
(\ref{cuatro}) does not modify the action]. Yet, there is another
approach, which does not employs vector potentials, that derives
charge quantization from the consistency of the Heisenberg
equations of motion of a charged particle in the field of a
magnetic monopole \cite{jackiw-monopolo}. But in the absence of
electric charges, all this is automatically guaranteed, and
Dirac's quantization condition is not required to have a
consistent theory.

It would be interesting, in view of the above discussion, to study
the LR formulation of the theory in the case with both charges and
monopoles. Also, the present approach could  be generalized to the
study of higher-rank Abelian theories, with their corresponding
extended objects generalizing electric and magnetic charges
\cite{deser}.


\begin{thebibliography}{25}


\bibitem{dirac}
P.A. M Dirac, Proc.\ Roy.\ Soc.\ Lond. {\bf A 113}, 60, (1931)
\bibitem{dirac2}
P.A.M. Dirac, Phys.\ Rev.\ {\bf 74 }, 317, (1948)

\bibitem{dibartolo}
C. di Bartolo, F. Nori, R. Gambini and A. Tr\'{i}as,\ Nuovo\
Cimento Soc. Ital. Fis. {\bf 38}, 497, (1983)

\bibitem{gambini1}
R. Gambini and A. Tr\'{i}as, Phys.\ Rev.\ D {\bf 27}, 2935, (1983)

\bibitem{gambini2}
R. Gambini and J. Pullin, Loops Knots, Gauge Theory and Quantum
Gravity (Cambridge University Press, 1996)

\bibitem{schwinger}
J. Schwinger, Phys.\ Rev.\ {\bf D12 }, 3105, (1975)

\bibitem{morandi}
G. Morandi, The Role of Topology in Classical and Quantum Physics
( Springer-Verlag, 1992)

\bibitem{balachandran}
A.P.Balachandran, G.Marmo, B.S.Skagerstam and A.Stern, Classical
Topology and Quantum States (World Scientific, 1991)

\bibitem{balachandran2}
A.P.Balachandran, Found. Phys. \textbf{24}, 455 (1994).

\bibitem{setaro}
R.Gambini and L.Setaro, Phys. Rev. Lett. \textbf{65}, 2623 (1990)

\bibitem{leal-ab}
L.Leal, Mod. Phys. Lett.\textbf{A9}, 1945 (1994)

\bibitem{dirac3}
P.A.M. Dirac, Lectures on quantum mechanics (Yeshiva University,
New York,1964)


\bibitem{jackiw}
R. Jackiw, ``\emph{Constrained quantization without tears}'', at
2nd. Workshop on Constraint Theory and Quantization Methods,
Montepulciano, Italy (1993) (hep-th/9306075).

\bibitem{camacaro}
J.Camacaro, R.Gaitan and L.Leal, \ Mod. \ Phys.\ Lett.\ {\bf A12
},3081, (1997)

\bibitem{pio1} P.~J.~Arias, C.~Di Bartolo, X.~Fustero, R.~Gambini and A.~Tr\'{i}as,
Int.\ J.\ Mod.\ Phys.\ {\bf A7}, 737 (1992)

\bibitem{pio}
P.J.Arias, E.Fuenmayor and L.Leal, Phys.\ Rev.\ D {\bf 69 },
125010 (2004)

\bibitem{deser}
S Deser, A Gomberoff, M.Henneaux, C. Teitelboin, Nucl.\ Phys.\
{\bf B520}, 179 (1998).


\bibitem{jackiw-monopolo}
R. Jackiw , ``\emph{Dirac's Magnetic Monopoles (Again)}'', at
Dirac Memorial Symposium, Tallahassee, Florida, (2002)
(hep-th/0212058).








\end{thebibliography}
\end{document}